# On the Selection of Tuning Methodology of FOPID Controllers for the Control of Higher Order Processes

Saptarshi Das[1,2], Suman Saha[1,3], Shantanu Das[4], Amitava Gupta[1,2]

1. School of Nuclear Studies & Applications, Jadavpur University, Salt Lake Campus, LB-8, Sector 3, Kolkata-700098, India. Email: saptarshi@pe.jusl.ac.in

2. Department of Power Engineering, Jadavpur University, Salt Lake Campus, LB-8, Sector 3, Kolkata-700098, India. Email: amitg@pe.jusl.ac.in

3. Drives and Control System Technology Group, Central Mechanical Engineering Research Institute, Mahatma Gandhi Avenue, Durgapur-713209, India. Email: s_saha@cmeri.res.in

4. Reactor Control Division, Bhabha Atomic Research Centre, Mumbai-400085, India. Email: shantanu@magnum.barc.gov.in

**Abstract**

In this paper, a comparative study is done on the time and frequency domain tuning strategies for fractional order (FO) PID controllers to handle higher order processes. A new fractional order template for reduced parameter modeling of stable minimum/non-minimum phase higher order processes is introduced and its advantage in frequency domain tuning of FOPID controllers is also presented. The time domain optimal tuning of FOPID controllers have also been carried out to handle these higher order processes by performing optimization with various integral performance indices. The paper highlights on the practical control system implementation issues like flexibility of online autotuning, reduced control signal and actuator size, capability of measurement noise filtration, load disturbance suppression, robustness against parameter uncertainties etc. in light of the above tuning methodologies.

**Keywords:** Fractional order controller, FOPID tuning, integral performance indices, NIOPTD, optimal time domain tuning, robust frequency domain tuning.

## 1. Introduction:

Modelling of process plants for control analysis and design often give rise to higher order models in order to capture delicate dynamic behaviours of the process, with higher accuracy [1]-[3]. It has been shown by Saha *et al.* [4] that a nonlinear process dynamics under shift in operating point can be nicely captured using system identification techniques with several higher order process models and then a generalized varying gain model. Tuning of suitable controllers for these higher order processes are a bit challenging. It is well known that among various types of industrial controllers, PID dominates most of the process control applications due to its simple structure, easy tuning and robustness [5], [6]. In recent past, FOPID or $PI^{\lambda}D^{\mu}$ controllers have been proposed by Podlubny [7] which are capable of enhancing the closed loop performance of a system over a simple integer order PID structure [8], [9]. In fact, the true potential of a FOPID controller greatly depends on its tuning methodology and often the performance may degrade severely, with contradictory design specifications to be met by the FO controllers. The present work attempts to show the inherent advantages and limitations of different tuning strategies, while designing FOPID controllers for higher order processes with specified time/frequency response.

In [5], [6], it has been shown that a reduced order model is required for a higher order plant before its tuning with a PID controller using classical tuning rules. Classical model reduction techniques for PID tuning mostly used First Order Plus Time Delay (FOPTD) and Second Order Plus Time Delay (SOPTD) templates, which are enhanced in this paper, with the introduction of new templates known as Non-Integer Order Plus Time Delay (NIOPTD) having *flexible order* elements. This allows robust iso-damped tuning of FOPID controllers without compromising the accuracy of the reduced order models. In other words, with the introduction of NIOPTD templates, robustness of a FOPID controller can be increased by the significant reduction in modelling error. In conventional frequency domain model reduction the suboptimal approach, proposed by Xue *et al.* [10] by minimizing the $H_2$ norm between the reduced order and the higher order process are popular among research communities and is also capable of extracting the delays in a model which finds great scope of application especially in building reduced order process models.

For process control applications, FO controllers have been classified in four categories in [11] among which Podlubny's $PI^\lambda D^\mu$ or FOPID [7] and Oustaloup's CRONE controller [12] and its three generations [13]-[15] deserve special merit. Other FO controllers like the FO lead-lag compensator [16] and FO phase shaper [17]-[19], [4] are also becoming popular in recent robust process control applications. Several tuning strategies have been proposed by many contemporary researchers to tune FOPID controllers in both frequency domain and time domain. It has been found that the frequency domain design technique requires a reduced order template of the original higher order process. The time domain tuning techniques, on the other hand, do not necessarily require a reduced order model and hence the higher order process model is sufficient to find out the controller parameters by an optimization technique with some time domain performance indices as the design criteria. In present day, most of the industrial controllers are tuned with a few set of design specifications, either in time domain (e.g. error index, rise time, percentage of overshoot, settling time, overshoot-undershoot ratio etc.) or frequency domain (e.g. gain margin, phase margin, cross-over frequencies, maximum sensitivity and complementary sensitivity magnitudes etc.) [5], [6]. Hence, a single tuning methodology can not satisfy all of the above design criteria i.e. simultaneously satisfying time and frequency domain performance specifications. Indeed, such contradictory design criteria may often give unsatisfactory, even unstable closed loop response due to over-specification. Thus, a FOPID controller, as argued above, satisfying few set of time domain specifications may not have sufficient robustness against system parameter uncertainties in frequency domain analysis and vice versa. Thus, it is clear, that every tuning strategy possesses its own inherent strength and weakness. The present work tries to focus on those characteristics of some well-established tuning approaches and their extensions for FOPID controllers in a comparative manner.

This paper also proposes a robust frequency domain tuning strategy FOPID controllers using highly accurate NIOPTD-II template for open loop stable, minimum/non-minimum phase higher order processes. The proposed technique uses a simultaneous nonlinear equation solving based robust tuning of FOPID controllers, which requires lesser computational load unlike a constrained nonlinear optimization used by the contemporary researchers for iso-damped tuning of FOPIDs. Further, it is seen that time domain optimal tuning method for FOPIDs, as in recent literaures, do not always guarantee the closed loop stability of the process. An extension of FOPID tuning strategy is proposed in this paper to guarantee the closed loop stability and also to select the most suitable integral performance index for optimal time domain tuning.

The rest of the paper is organized as follows. Tuning methodologies for FOPID controllers, proposed by contemporary researchers are outlined in section 2. In section 3, the robust frequency domain tuning of FOPID controller is discussed along with the proposal for new model reduction templates and simulations with a test-bench of higher order processes. Section 4 discusses about the time domain optimal FOPID design by minimizing a chosen time domain integral performance index. In section 5, the design performances of the robust and optimal FOPID controllers are compared from various perspectives of control system analysis. The paper ends with the contributions of the present work as the conclusion in section 6, followed by the references.

## 2. Tuning of FOPID controllers: review of the existing methodologies:

Several methods have been proposed for tuning $PI^\lambda D^\mu$ controllers [7] by many contemporary researchers. A Ziegler-Nichols type empirical rule for tuning of $PI^\lambda D^\mu$ controllers has been proposed by Valerio & Sa da Costa [20]. Fractional MIGO [21] based tuning rule for FOPI controllers has been developed by Chen *et al.* [22]. But these methodologies [20], [22] need the reduced order models of a higher order process to take a FOPTD form only, which may not be sufficient to describe the complex dynamic behaviour of the higher order plant as discussed earlier and addressed in the present work with highly accurate NIOPTD templates. From specified phase margin ($\phi_m$), gain crossover frequency ($\omega_{gc}$) [23] and iso-damping/robustness criteria (i.e. flat phase curve around $\omega_{gc}$) [24], [25] a tuning methodology for FOPI/FOPD controllers for controlling integer order systems have been discussed in [26], [27]. The same tuning strategy for a class of fractional order models have been extended by Luo *et al.* [28], [29]. An optimization based frequency domain tuning method for $PI^\lambda D^\mu$ controller has been proposed by Monje *et al.* [30] and Dorcak *et al.* [31] which takes two extra specifications on maximum value of the magnitude of sensitivity and complementary sensitivity function along with the specifications presented in [27]-[29]. Time domain techniques of FOPID controller tuning includes dominant pole placement tuning [32]-[33] and optimal tuning [34]-[37] based on time domain integral performance index [38] minimization. The dominant pole placement tuning, proposed by Biswas *et al.* [32] and Maiti *et al.* [33] is only valid for strictly second order type systems and it does not give satisfactory result for higher order systems having several dominant poles and/or zeros. Also, the dominant pole placement tuning [32], [33] gives inferior closed loop performance and often unstable response for time delay systems, since the Pade approximation of delay term effectively raises the order of the overall system. The time domain optimal tuning method proposed by Cao & Cao [34], [35] and Maiti *et al.* [36] often fails to guarantee the close loop stability of the process along with the FOPID controller.

Zamani *et al.* [39] proposed a stochastic optimization based tuning with a customized cost function consisting of various control objectives like maximum overshoot, rise time, settling time, steady state error, Integral of Absolute Error (IAE), squared control signal, inverse of phase margin and gain margin. Alomoush [37] optimized Integral of Time multiplied Absolute Error (ITAE) and Lee & Chang [40] optimized Integral of Square Error (ISE) as the integral performance index to find out the optimal set of controller parameters. An optimization based controller tuning by minimizing matrix norms as the cost functions has been proposed by Bouafoura & Braiek [41]. Castillo *et al.* [42] proposed a tuning methodology for FOPI controllers for first order systems from frequency domain specifications while also meeting few set of time domain specifications simultaneously. Bhambhani *et al.* [43] proposed a multi-objective optimization based FOPI controller tuning methodology for Networked Control Systems (NCS) which simultaneously minimizes ITAE of the closed loop system and maximizes the jitter margin. Thus it can be seen that several time domain integral performance indices have been optimized by many contemporary researchers. Tavazoei in [38] has given a brief description of the finiteness of the integral performance indices for fractional order systems for step input and load disturbance excitation, which is required to be taken into account before the optimization. Caponetto *et al.* [44] investigated stabilization of FOPTD processes with FOPID controllers. A similar stabilization problem with FOPD/FOPID controller for integer order integrating processes has been discussed by Hamamci & Koksal [45] and fractional order integrating processes by Hamamci [46]. A $PI^{\lambda}D^{\mu}$ controller design for FO systems based on extended root-locus method has also been studied by Bayat & Ghartemani [47]. Recently, Padula & Visioli [48] proposed empirical tuning rules for FOPID controllers using IAE minimization criteria with constraints on maximum sensitivity for the FOPTD processes, which is rather a simplified approximation for higher order processes with large modeling error.

The present approach automatically takes care of the stability of the closed loop system while tuning the FOPID controller in time domain. Also, the frequency domain controller tuning techniques [23] are applied to FOPID controllers for NIOPTD reduced order models, which require no optimization (deterministic [26], [30], [37] or stochastic [31]-[36], [39], [40]) but a simultaneous nonlinear equation solving technique and hence lesser computational load. The novelty of the work with respect to the available techniques is to formulate a FOPID tuning stategy for the control of higher order processes in two different ways i.e. frequency domain and time domain approach and also highlighting the inadequacies inherent in these tuning philosophies.

## 3. Frequency domain design of FOPID controllers
### 3.1. Frequency domain design specifications for robust FOPID tuning

Frequency domain design [23] of FOPID controllers was proposed by Monje *et al.* [30] based on a constrained optimization problem.

i.e. If $P(s)$ be the model of the process plant, then the objective is to find a controller $C(s)$, so that the open loop system $G(s) = C(s)P(s)$ would meet the following design specifications:

(a) Phase margin specification:

$$Arg[G(j\omega_{gc})] = Arg[C(j\omega_{gc})P(j\omega_{gc})]$$
$$= -\pi + \phi_m \tag{1}$$

(b) Gain crossover frequency specification:

$$\left| G(j\omega_{gc}) \right| = \left| C(j\omega_{gc})P(j\omega_{gc}) \right| = 1 \tag{2}$$

(c) Robustness to gain variation (Iso-damping):

$$\left( \frac{d}{d\omega} \left( Arg[G(j\omega)] \right) \right)_{\omega=\omega_{gc}} = 0 \tag{3}$$

(d) Complementary sensitivity specification:

$$\left| T(j\omega) \right| = \left| \frac{C(j\omega)P(j\omega)}{1+C(j\omega)P(j\omega)} \right|_{dB} \leq A\, dB \quad \forall \omega \geq \omega_t\, rad/s \tag{4}$$

$$\Rightarrow \left| T(j\omega_t) \right| = A\, dB$$

where, $A$ is the specified magnitude of the complementary sensitivity function or noise attenuation for frequencies $\omega \geq \omega_t\, rad/s$.

(e) Sensitivity specification:

$$\left| S(j\omega) \right| = \left| \frac{1}{1+C(j\omega)P(j\omega)} \right|_{dB} \leq B\, dB \quad \forall \omega \leq \omega_s\, rad/s \tag{5}$$

$$\Rightarrow \left| S(j\omega_s) \right| = B\, dB$$

where, $B$ is the specified magnitude of the sensitivity function or load disturbance suppression for frequencies $\omega \leq \omega_s\, rad/s$.

(f) Elimination of Steady-state error:

The steady-state error of the closed loop system automatically gets cancelled with the introduction of the fractional integrator.

Clearly, while designing FO-controllers with the specifications (1)-(5), it is first required to know the frequency response of the higher order plants i.e. $P(j\omega)$ in an accurate reduced order template and also of the chosen controller structure i.e. $C(j\omega)$. Monje *et al.* [26], [30] has reported the results of tuning simple FOPTD plants with FOPID controllers. Indeed, the above methodology can not be directly applied to tune any arbitrary higher order process model without reducing it in prespecified structure. Hence, the chosen reduced parameter structure should be flexible enough to capture large variety of arbitrary higher order models with high accuracy since modeling inaccuracy with FOPTD and SOPTD structures might reduce the achievable robustness of a FOPID controller. In the next subsection, the new reduced parameter templates are introduced which have higher capability of retaining the dominant dynamics of higher order models than the classically used FOPTD and SOPTD structures.

### 3.2. New approach towards reduced parameter FO modeling of higher order processes

In conventional process control applications higher order process models are approximated using FOPTD and SOPTD structures given by:

(a) First Order Plus Time Delay (FOPTD):

$$P^I(s) = \frac{K}{Ts+1}e^{-Ls} \tag{6}$$

(b) Second Order Plus Time Delay (SOPTD):

$$P^{II}(s) = \frac{K}{s^2 + 2\zeta\omega_n s + \omega_n^2}e^{-Ls} \tag{7}$$

For higher order linear models these structures give large modelling error and this proves the inadequacy of model reduction with FOPTD and SOPTD template for robust FOPID design. Hence, to obtain better accuracy of the reduced order models, two new structures, involving FO elements, have been proposed here. The noninteger reduced parameter models are defined as:

(c) One Non-integer Order Plus Time Delay (NIOPTD-I):

$$P^{III}(s) = \frac{K}{Ts^\alpha + 1}e^{-Ls} \tag{8}$$

(d) Two Non-integer Orders Plus Time Delay (NIOPTD-II):

$$P^{IV}(s) = \frac{K}{s^\alpha + 2\zeta\omega_n s^\beta + \omega_n^2}e^{-Ls} \tag{9}$$

Here, the system parameters have their classical meanings and the additional two parameters i.e. the system orders $\{\alpha, \beta\}$ are allowed to take any real value and hence can be termed as *flexible orders* of the compressed models.

Now, the model compression of higher order processes are formulated with the help of an optimization based technique. Let us consider any arbitrary higher order minimum/non-minimum phase stable transfer function $P(s)$ which is to be modelled as a reduced order one $\tilde{P}(s)$ having *flexible order* elements. The frequency domain performance index ($J_f$) for model reduction is taken as the deviation of $H_2$ norm of the original and reduced systems as studied in integer order domain by Xue *et al.* [10]. The $H_2$ norm of a system reflects how much it amplifies or attenuates its inputs over all the frequencies. In other words, it represents the energy of the output signal of a system, subjected to an impulse exciation. Mathematically, $H_2$ norm of a system $P(s)$ can be evaluated by the following relation

$$\|P(s)\|_2 = \sqrt{\frac{1}{2\pi}\int_{-\infty}^{\infty} trace\left[P(j\omega)\overline{P(j\omega)^T}\right]d\omega} \tag{10}$$

Here, $J_f = \|P(s) - \tilde{P}(s)\|_2$ \hfill (11)

During the optimization process, each guess values involving arbitrary fractional order elements are rationalized by a fifth order Oustaloup's approximation within the frequency range $\omega \in [10^{-4}, 10^4]$ rad/s. The performance index $J_f$ (11) is then minimized with unconstrained Nelder-Mead Simplex algorithm [49] implemented in MATLAB's Optimization Toolbox [50] function *fminsearch()* to obtain a suitable set of values of reduced order model parameters i.e. $\{K, T, \alpha, L\}$ for NIOPTD-I and $\{K, \zeta, \omega_n, \alpha, \beta, L\}$ for NIOPTD-II structure. It is clear that FOPTD and SOPTD are just the special cases of the two new proposed templates (NIOPTD-I and NIOPTD-II respectively). In FOPTD and SOPTD modelling the order of the compressed models are forced to take integer values only which essentially means only model parameters are to be searched and not the corresponding orders.

The above model compression technique is now applied to the the following higher order test bench process plants (1)-(4) as presented in Astrom *et al.* [51], Panagopoulos *et al.* [52], Shen [53] and Chen *et al.* [22]

$$P_1(s) = \frac{1}{(s+1)^3} \tag{12}$$

$$P_2(s) = \frac{9}{(s+1)(s^2+2s+9)} \tag{13}$$

$$P_3(s) = \frac{1}{(s+1)^4} \tag{14}$$

$$P_4(s) = \frac{1}{(s+1)(0.2s+1)(0.04s+1)(0.008s+1)} \tag{15}$$

Table 1
Choice of reduced parameter model structure based on minimum modeling error

| Process model | Minima of the objective function for reduced parameter models with different structures | | | | Preferred structure |
|---|---|---|---|---|---|
| | FOPTD | SOPTD | NIOPTD-I | NIOPTD-II | |
| $P_1$ | 0.6961 | 0.0859 | 0.5477 | 0.0434 | NIOPTD-II |
| $P_2$ | 0.6132 | 0.4303 | 0.6129 | 0.1006 | NIOPTD-II |
| $P_3$ | 0.8480 | 0.1505 | 0.6568 | 0.0893 | NIOPTD-II |
| $P_4$ | 0.2138 | 0.0137 | 0.1960 | 0.0099 | NIOPTD-II |

Now, the most suitable structure for model reduction can be decided from the minimum value of the objective function (11) or modelling error by an optimization with structures (6)-(9). The optimization results with the above process models are presented in Table 1. It is evident from Table 1, that optimization with the proposed NIOPTD-II structure leads to a better minimization of the modelling error than that with the other ones. The corresponding reduced parameter models are reported in Table 2.

It is also found that each of the reduced order models have a delay term, whereas the original plant transfer function was delay-free. This can be justified from the fact that in most of the process plants, the initial rate of rise of the transient response is slow due to its inherent inertia, and then gradually the transient response tracks the input excitation and finally settles down. Thus, an apparent delay has been estimated to make the process modelling more realistic, as reported in [3], [5].

Table 2
Reduced parameter models of the test-bench process plants.

| Process model | Class of Models | Reduced order Models | |
|---|---|---|---|
| | | FOPTD or NIOPTD-I | SOPTD or NIOPTD-II |
| $P_1$ | Integer Order | $\dfrac{1}{2.029s+1}e^{-1.1349s}$ | $\dfrac{0.50679}{s^2+1.3177s+0.50679}e^{-0.4009s}$ |
| | Fractional Order | $\dfrac{0.99391}{2.3298s^{1.0648}+1}e^{-1.0006s}$ | $\dfrac{0.42456}{s^{2.109}+1.2157s^{1.015}+0.42515}e^{-0.2694s}$ |
| $P_2$ | Integer Order | $\dfrac{1}{0.88889s+1}e^{-0.4149s}$ | $\dfrac{5.3871}{s^2+4.7886s+5.3871}e^{-0.2650s}$ |
| | Fractional Order | $\dfrac{1.0003}{0.8864s^{1.0212}+1}e^{-0.4274s}$ | $\dfrac{4.4659}{s^{2.4673}+5.2284s^{1.0201}+4.4701}e^{-0.1217s}$ |
| $P_3$ | Integer Order | $\dfrac{1.0001}{2.3456s+1}e^{-1.8713s}$ | $\dfrac{0.34247}{s^2+1.0512s+0.34247}e^{-0.9361s}$ |
| | Fractional Order | $\dfrac{0.99149}{2.8015s^{1.0759}+1}e^{-1.6745s}$ | $\dfrac{0.22287}{s^{2.2251}+0.86316s^{1.0389}+0.22394}e^{-0.5532s}$ |
| $P_4$ | Integer Order | $\dfrac{1}{1.0564s+1}e^{-0.2097s}$ | $\dfrac{4.6812}{s^2+5.6676s+4.6812}e^{-0.0421s}$ |
| | Fractional Order | $\dfrac{0.99932}{1.0842s^{1.0132}+1}e^{-0.1922s}$ | $\dfrac{5.069}{s^{1.9954}+6.0645s^{0.99973}+5.069}e^{-0.0518s}$ |

### 3.3. Tuning results of FOPID controllers based on NIOPTD-II models

The robust frequency domain design of FOPID controllers was first proposed by Monje *et al.* [26], [30] and Dorcak *et al.* [31], based on a constrained nonlinear optimization with frequency domain specifications. An analytical method with simultaneous equation solving to estimate the FO-controller parameters for first order and one noninteger order class of models can be found in Li *et al.* [27] and Luo *et al.* [28], [29]. Here, the controller designing methodology is carried out with the most accurate NIOPTD-II reduced order models in Table 2 for the test plants (12)-(15). The structure of the FOPID controller considered here is in the parallel/noninteracting form

$$C(s)=K_p+\frac{K_i}{s^\lambda}+K_d s^\mu \tag{16}$$

The frequency domain tuning with the specifications (1)-(5) basically uses the gain, phase and phase derivative which is now derived for the reduced parameter NIOPTD-II model and FOPID controller. The gain and phase of the NIOPTD-II structure (9) is given by

$$|P(j\omega)|=\frac{K}{\sqrt{\left(\omega^\alpha\cos\dfrac{\alpha\pi}{2}+2\zeta\omega_n\omega^\beta\cos\dfrac{\beta\pi}{2}+\omega_n^2\right)^2+\left(\omega^\alpha\sin\dfrac{\alpha\pi}{2}+2\zeta\omega_n\omega^\beta\sin\dfrac{\beta\pi}{2}\right)^2}} \tag{17}$$

$$Arg[P(j\omega)] = -\omega L - \tan^{-1}\left( \frac{\omega^\alpha \sin\frac{\alpha\pi}{2} + 2\zeta\omega_n\omega^\beta \sin\frac{\beta\pi}{2}}{\omega^\alpha \cos\frac{\alpha\pi}{2} + 2\zeta\omega_n\omega^\beta \cos\frac{\beta\pi}{2} + \omega_n^2} \right) \tag{18}$$

Also, the derivative of phase of the model (9) with respect to frequency ($\omega$) is

$$\left( \frac{d}{d\omega}\big(Arg[P(j\omega)]\big) \right) = -L - \frac{\left( 2\zeta\omega_n(\alpha-\beta)\,\omega^{\alpha+\beta-1}\sin\frac{(\alpha-\beta)\pi}{2} \right) - \left( \alpha\omega_n^2\omega^{\alpha-1}\sin\frac{\alpha\pi}{2} \right) - \left( 2\beta\zeta\omega_n^3\omega^{\beta-1}\sin\frac{\beta\pi}{2} \right)}{\left( \omega^\alpha \cos\frac{\alpha\pi}{2} + 2\zeta\omega_n\omega^\beta \cos\frac{\beta\pi}{2} + \omega_n^2 \right)^2 + \left( \omega^\alpha \sin\frac{\alpha\pi}{2} + 2\zeta\omega_n\omega^\beta \sin\frac{\beta\pi}{2} \right)^2}$$

$$\tag{19}$$

The gain and phase of the FOPID controller (16) is given as

$$\left| C(j\omega) \right| = \sqrt{\left( K_p + K_i\omega^{-\lambda}\cos\frac{\lambda\pi}{2} + K_d\omega^\mu\cos\frac{\mu\pi}{2} \right)^2 + \left( K_d\omega^\mu\sin\frac{\mu\pi}{2} - K_i\omega^{-\lambda}\cos\frac{\lambda\pi}{2} \right)^2} \tag{20}$$

$$Arg\left[ C(j\omega) \right] = \tan^{-1}\left( \frac{K_d\omega^\mu\sin\frac{\mu\pi}{2} - K_i\omega^{-\lambda}\cos\frac{\lambda\pi}{2}}{K_p + K_i\omega^{-\lambda}\cos\frac{\lambda\pi}{2} + K_d\omega^\mu\cos\frac{\mu\pi}{2}} \right) \tag{21}$$

The derivative of phase of the controller (16) with respect to frequency ($\omega$) is

$$\left( \frac{d}{d\omega}\big(Arg[C(j\omega)]\big) \right) = \frac{\left( K_pK_d\mu\omega^{\mu-1}\sin\frac{\mu\pi}{2} \right) + \left( K_pK_i\lambda\omega^{-\lambda-1}\sin\frac{\lambda\pi}{2} \right) + \left( K_iK_d(\lambda+\mu)\omega^{-\lambda+\mu-1}\sin\frac{(\lambda+\mu)\pi}{2} \right)}{\left( K_p + K_i\omega^{-\lambda}\cos\frac{\lambda\pi}{2} + K_d\omega^\mu\cos\frac{\mu\pi}{2} \right)^2 + \left( K_d\omega^\mu\sin\frac{\mu\pi}{2} - K_i\omega^{-\lambda}\cos\frac{\lambda\pi}{2} \right)^2}$$

$$\tag{22}$$

Now, having known the frequency response of the reduced NIOPTD-II models (9) and FOPID controllers (16), by satisfying the design specifications (1)-(5), the controller parameters can be calculated. The FOPID controller (16) has five parameters to tune, i.e. $\{K_p, K_i, K_d, \lambda, \mu\}$ which can be found out with the five design specifications (1)-(5). In Luo *et al.* [28], [29] and Li *et al.* [27], it has been shown that from each expression of gain, phase and derivative of phase for the controller and plant, the controller parameters can be determined analytically or graphically. In the present work, all the model parameters of the highly accurate NIOPTD-II structure (9) i.e. pseudo-dc gain ($K$), damping ratio ($\zeta$), undamped natural frequency ($\omega_n$), two-dominant fractional orders of the system ($\alpha, \beta$) and transport delay ($L$) have been evaluated in Table 2. Also, from the design specifications (1)-(5), the desired phase margin ($\phi_m$) and gain crossover frequency ($\omega_{gc}$), sensitivity and complementary sensitivity magnitudes ($\left| S(j\omega_s) \right|$ and $\left| T(j\omega_t) \right|$) are known. So with these known values, the controller parameters $\{K_p, K_i, K_d, \lambda, \mu\}$ can be solved out from equation (1)-(5). But the problem is that an explicit analytical solution is not so easy to derive when the controller and model structure itself are much complicated. Also, depending on a fixed model, a predefined graphical solution [27]-[29] restricts the application from the flexibility of online auto-tuning of the controller parameters. It is also observed that gain and phase equations for the model and controller both are implicit in nature containing nonlinear and transcendental terms. So, simple analytical or even classical simultaneous linear equation solving techniques can not be applied in this case to solve out the controller parameters from equation (1)-(5). As a solution to this, Powell's Trust-Region-Dogleg algorithm, implemented in MATLAB's Optimization Toolbox [50] function *fsolve()* is used to find out the value of the controller parameters $\{K_p, K_i, K_d, \lambda, \mu\}$. Function *fsolve()* is capable of estimating the numerical solution of simultaneous nonlinear or transcendental equations. In Table 1, it has been shown that compared to other reduced order structures, the NIOPTD-II can capture the higher order dynamics of a process model much efficiently and hence in the present study only the accurate NIOPTD-II model structure is used for the frequency domain tuning of FOPID controllers. Now, For FOPID controller tuning maximum magnitude of complementary sensitivity and sensitivity functions have been selected as $A = -40dB$ and $B = -40dB$ at $\omega_t = 10 \, rad/\sec$ and $\omega_s = 10^{-2} \, rad/\sec$ respectively. Other tuning specifications are similar to that of Luo's [28], [29], Li's [27] and Monje's [26], [30] works. The tuned controller parameters for all of the reduced parameter NIOPTD-II models of the respective test plants are given in Table 3 along with the phase margin ($\phi_m$) and gain-cross over frequency ($\omega_{gc}$) specifications that have been used for tuning. It has been also observed that design with the nonlinear simultaneous equation solver *fsolve()* converges in most of the cases, whereas the same formulation may not converge with constrained optimization solver *fmincon()* as proposed by Monje *et al.* [26], [30]. Numerical solution with function *fsolve()* may diverge for simultaneous demand of large $\phi_m$ for low overshoot and also demand of high $\omega_{gc}$ to get faster time response. In such cases, the designer should initially tune the plant at lower $\omega_{gc}$ similar to that presented in [26]-[30]. Now, with a sufficiently large flat phase curve around $\omega_{gc}$, system's dc gain can be increased to get faster time response by keeping the overshoot at the same level. The objective of iso-damped frequency domain tuning for the family of FOPID controllers, presented in this section, is to achieve gain independent overshoot in some specific robust control applications like Saha *et al.* [4] and Chao *et al.* [54].

Table 3: Frequency domain tuning results of FOPID controllers for test-bench processes

| Process | Design Specifications | | FOPID Controller Parameters | | | | |
|---------|---------------------|--------------------|---------|---------|---------|---------|---------|
| | $\phi_m$ **(degree)** | $\omega_{gc}$ **(rad/s)** | $K_p$ | $K_i$ | $K_d$ | $\lambda$ | $\mu$ |
| $P_1$ | 80 | 0.3 | 0.9116 | 0.2526 | 0.2023 | 1.1577 | 0.9973 |
| $P_2$ | 80 | 1.0 | 0.8444 | 1.2309 | 0.2713 | 1.0019 | 0.9355 |
| $P_3$ | 80 | 0.1 | 0.3677 | 0.0781 | 0.0992 | 1.1204 | 1.0158 |
| $P_4$ | 80 | 1.0 | 0.9007 | 1.3198 | 0.3196 | 0.9495 | 0.9284 |

The corresponding Bode diagram (Fig. 1) shows wide flatness in the phase curves around the gain cross-over frequencies which ensures iso-damped time responses (Fig. 2).

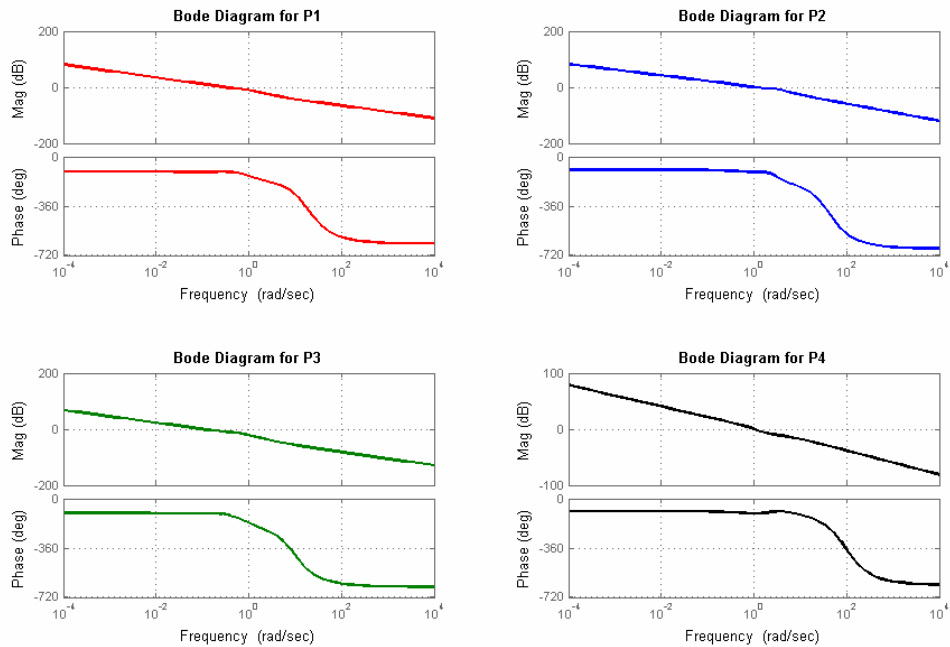

Fig. 1. Bode diagram of the open loop system comprising of the NIOPTD-II model and robust FOPID controller.

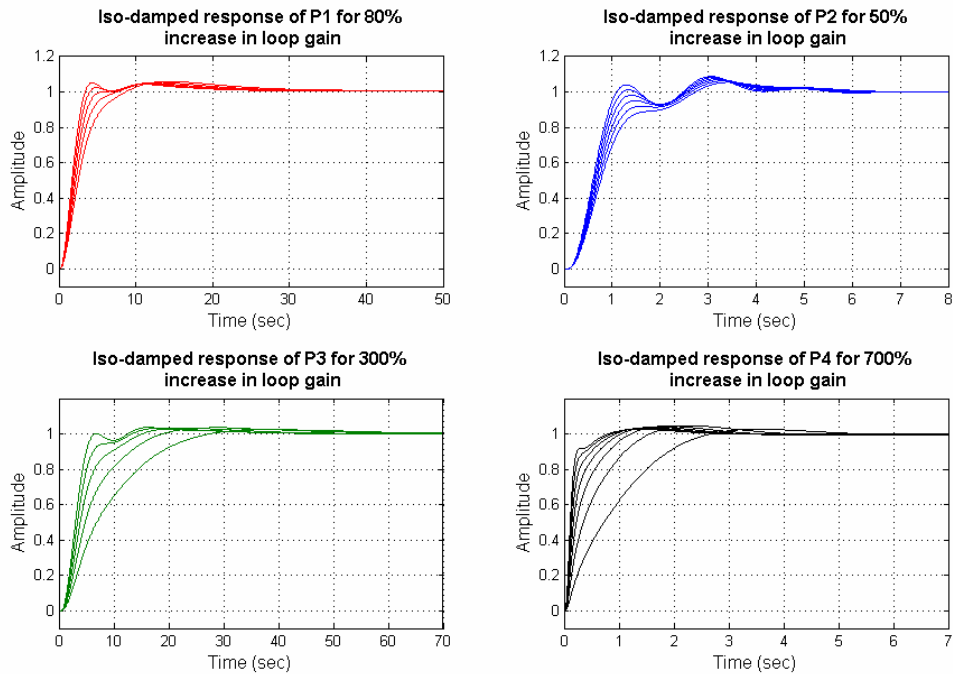

Fig. 2. Iso-damped closed loop response of the test-bench plants.

From Fig. 2 it is evident that the frequency domain design allows high level of loop gain variation which not only ensures good time domain performance under modeling uncertainties but also makes the system faster for increase in loop gain while maintaining the overshoot at same level. In all the above studies, the delay terms in the reduced order process models are simulated with third order Pade approximation.

## 4. Time domain design of FOPID controllers

In this section, the time domain optimal tuning method of FOPID controllers has been formulated for the control of higher order processes (12)-(15). This technique searches for an optimal set of controller parameters while minimizing a suitable time domain integral performance index [6], [38]. This methodology of FOPID controller synthesis does not require any model reduction in a generalized template of the actual higher order process, since time domain performance indices can be evaluated from the original/identified process model directly, provided the knowledge of the process model is fairly accurate from its governing physical laws or classical identification techniques.

Also, the time domain optimization based tuning methodology can not be applied directly without restricting the unstable modes of the closed loop system within the search space. The controller parameters $\{K_p, K_i, K_d, \lambda, \mu\}$ are searched while minimizing a suitable time domain performance index (as the objective function) such that the closed loop system comprising of the open loop plant along with the FOPID controller be stable and finite with the help of MATLAB's *isstable()* and *isfinite()* functions respectively.

As discussed in section 2, another time domain tuning method, known as dominant pole placement tuning of FOPID controllers proposed by Biswas *et al.* [32], Maiti *et al.* [33] can not guarntee the closed loop stability of the system in the optimization process. Also it gives inferior time response for higher order or time delay systems since the concept is based on the assumption that the dynamics of any arbitrary higher order system is dominated by two complex conjugate poles, which may not be the reality from controller designing point of view, especially for high gain and long delay systems like [4] and also systems with several dominant poles/zeros. Strictly second order systems with no delay, theoretically can be controlled by dominant pole placement technique and it has been found that performance is not satisfactory for systems with large time-delay, higher order and also fractional order systems, having several dominant poles and zeros. So, the present study is restricted in the optimal time domain performance index based tuning only for performance study of FOPID controllers.

### 4.1. Choice of a suitable time domain integral performance index:

It is well known that if the control loop error be $e(t)$, the control signal of a FOPID controller [7], [24] is given by

$$u(t) = K_p e(t) + K_i D^{-\lambda} \left[ e(t) \right] + K_d D^{\mu} \left[ e(t) \right] \tag{23}$$

where, "$D$" signifies the differential operator of fractional order and the negative power of $D$ signifies an integral action of fractional order [24].

Now, the simple error minimization criteria can be customized by a suitable choice of a time domain performance index (PI) to have a better control action as reported in [6], [34]-[40], [55] i.e.

(a) Integral of Absolute Error (IAE):

$$IAE = \int_0^\infty |e(t)| dt \tag{24}$$

(b) Integral of Time multiplied Absolute Error (ITAE):

$$ITAE = \int_0^\infty t \left| e(t) \right| dt \tag{25}$$

(c) Integral of Square Error (ISE):

$$ISE = \int_0^\infty e^2(t) \, dt \tag{26}$$

(d) Integral of Time multiplied Square Error (ITSE):

$$ITSE = \int_0^\infty t e^2(t) \, dt \tag{27}$$

(e) Integral of Squared Time multiplied by Error, all to be Squared (ISTES):

$$ISTES = \int_0^\infty \left[ t^2 e(t) \right]^2 dt \tag{28}$$

(f) Integral of Squared Time multiplied by Square Error (ISTSE):

$$ISTSE = \int_0^\infty t^2 e^2(t) \, dt \tag{29}$$

The presence of the time multiplication term and its higher powers in the performance indices (25), (27), (28), (29), puts more penalties on the chance of oscillation at later stages in the time response curve and thus effectively helps to reduce the settling time ($t_s$) of the closed loop system. Similarly, higher powers of error term put larger penalties for the larger values of $\left| e(t) \right|$ and thus minimize the chance of large overshoot. For practical implementation, the upper limits of the definite integrals in (24)-(29) are not possible to evaluate upto infinity and should be chosen to be sufficiently high so that the transient response decays out within that interval and the solution becomes independent of the choice of upper limit thereof.

Zamani *et al.* [39] proposed a customized performance index for optimization based tuning which minimizes sum of several specifications like overshoot, rise time, settling-time, steady-state error, absolute value of the error-signal, squared value of the controller outout signal and simultaneously maximizes the gain-margin and phase-margin. To show that a customized objective function comprising of several other performance indices like [39] indeed averages the true potential of each of them and deteriorates the performance of the closed loop system than each of the individual performance index, a new objective function has been formulated which is the sum of all the previous ones (24)-(29). The responses while minimizing each of the performance indices are also compared with the customized one (30) considering equal weights for all of its constituents. Putting the weights to zero in (30) except a single value indicates focussing towards a specific performance index based tuning.

$$J_{PI\_all} = w_1 \cdot IAE + w_2 \cdot ITAE + w_3 \cdot ISE + w_4 \cdot ITSE + w_5 \cdot ISTES + w_6 \cdot ISTSE \tag{30}$$

Cao & Cao [34], [35] and Maiti *et al.* [36] proposed a tuning strategy for fractional order controllers by minimizing the sum of IAE/ITAE and controller output. Whereas, the present work is attempted to extend the methodology for other available performance indices while also ensuring stability preserved convergence. In the present work, the performance indices (24)-(29) are evaluated using Trapezoidal rule for numerical integration and then minimized with the constrained Nelder-Mead Simplex algorithm [49] implemented in MATLAB's optimization toolbox [50] function *fmincon()* to obtain an optimal set of FOPID controller parameters. In this specific application the unconstrained optimization function *fminsearch()* should not be used, since the controller parameters (i.e. controller gains) may take very large values while searching for the minimum value of the objective functions, thus creating problem in practical implementation. In the present study, the controller parameters are searched within an interval $\{K_p, K_i, K_d\} \in [1, 500]; \{\lambda, \mu\} \in [0, 2]$ similar to that presented in [34]-[37]. Sometimes, MATLAB's constrained optimization function *fmincon()* may get trapped in local minimas. To ensure that the global minima has been found in the optimization process, the initial guesses of the controller parameters are perturbed enough and the simulation has been run several times and only the best results are reported. As discussed earlier, the optimal controller parameter search are restricted with the MATLAB functions *isstable()* and *isfinite()* to avoid the undesirable modes, especially the unstable modes. A large penalty function has been included in the objective function in each occurance of the undesirable modes which strongly discourages parameter search with unstable zones, as suggested by Zamani *et al.* [39].

### 4.2. Comparison of FOPID design with different performance indices

With the method as discussed in section 4.1, the parameters of $PI^\lambda D^\mu$ controller are now tuned for each test plants (12)-(15) and each time domain integral performance indices (24)-(29). The tuned parameters of the $PI^\lambda D^\mu$ controller for all of the test-bench process plants and corresponding closed loop performances i.e. the maximum percentage of overshoot ($\%M_p$) and rise-time ($t_r$) are reported in Table 4-7 respectively. The upper limit of the integral performance indices are chosen as 50 seconds. The corresponding closed loop responses are shown in Fig. 3-6.

Table 4:
Comparison of closed loop performance of plant $P_1$ with different performance indices.

| Performance Index | Minima of performance indices | $K_p$ | $K_i$ | $K_d$ | $\lambda$ | $\mu$ | $\%M_p$ | $t_r$ |
|---|---|---|---|---|---|---|---|---|
| IAE | 0.7761 | 6.5139 | 1.2006 | 5.1249 | 1.1538 | 1.3201 | 1.0 | 6.96 |
| ISE | 0.7279 | 2.6311 | 0.6586 | 4.0297 | 1.3116 | 0.6001 | 14.8 | 1.39 |
| ITAE | 5.2762 | 2.9692 | 0.8028 | 1.3394 | 1.0138 | 0.6074 | 23.1 | 1.78 |
| ITSE | 0.6950 | 2.7093 | 0.8476 | 1.4353 | 0.9103 | 0.8479 | 11.2 | 1.97 |
| ISTES | 98.9800 | 3.9007 | 1.4320 | 1.4485 | 1.053 | 0.8319 | 34.6 | 1.54 |
| ISTSE | 0.8153 | 2.2484 | 0.9553 | 1.1314 | 1.0094 | 0.8491 | 14.7 | 2.18 |
| Sum of all PIs ($J_{PI\_all}$) | 106.3568 | 3.3436 | 0.9177 | 1.2331 | 0.9774 | 0.9933 | 17.7 | 1.84 |

Table 5:

Comparison of closed loop performance of plant $P_2$ with different performance indices.

| Performance Index | Minima of performance indices | $K_p$ | $K_i$ | $K_d$ | $\lambda$ | $\mu$ | $\%M_p$ | $t_r$ |
|---|---|---|---|---|---|---|---|---|
| IAE | 1.1388 | 1.0891 | 1.0195 | 0.9577 | 0.9818 | 0.9786 | 1.1 | 5.05 |
| ISE | 0.2147 | 5.3242 | 1.4801 | 0.7641 | 1.5437 | 1.5606 | 7.1 | 2.67 |
| ITAE | 1.7622 | 1.1725 | 1.0461 | 1.0276 | 1.0008 | 0.8027 | 1.7 | 4.19 |
| ITSE | 0.1845 | 0.118 | 2.6198 | 1.7121 | 0.8415 | 0.7888 | 7.8 | 2.1 |
| ISTES | 1.5934 | 1.5652 | 1.2330 | 1.1167 | 0.9986 | 1.2333 | 2.2 | 3.6 |
| ISTSE | 0.5141 | 1.0671 | 1.0325 | 0.9836 | 1.0016 | 0.9472 | 2.4 | 4.59 |
| Sum of all PIs ($J_{PI\_all}$) | 10.1972 | 1.0273 | 0.9863 | 1.0110 | 1.0007 | 1.0058 | 2.5 | 4.58 |

Table 6:

Comparison of closed loop performance of plant $P_3$ with different performance indices.

| Performance Index | Minima of performance indices | $K_p$ | $K_i$ | $K_d$ | $\lambda$ | $\mu$ | $\%M_p$ | $t_r$ |
|---|---|---|---|---|---|---|---|---|
| IAE | 2.1936 | 1.4966 | 0.4696 | 1.4612 | 1.0000 | 1.034 | 1.9 | 4.39 |
| ISE | 0.3405 | 19.1897 | 4.2549 | 20.9892 | 0.8003 | 1.7214 | 20.4 | 0.64 |
| ITAE | 4.0269 | 1.2449 | 0.4220 | 1.1566 | 1.0000 | 0.9576 | 9.0 | 5.1 |
| ITSE | 0.5715 | 4.9274 | 0.6431 | 5.0121 | 1.0593 | 1.5743 | 7.4 | 3.08 |
| ISTES | 43.2518 | 1.1594 | 0.4688 | 1.1595 | 0.9994 | 0.8203 | 5.1 | 4.28 |
| ISTSE | 2.8172 | 1.1899 | 0.5029 | 1.3556 | 0.9831 | 0.8146 | 4.6 | 4.05 |
| Sum of all PIs ($J_{PI\_all}$) | 57.1613 | 1.1660 | 0.4730 | 1.1369 | 1.0000 | 0.8208 | 6.0 | 4.23 |

Table 7

Comparison of closed loop performance of plant $P_4$ with different performance indices.

| Performance Index | Minima of performance indices | $K_p$ | $K_i$ | $K_d$ | $\lambda$ | $\mu$ | $\%M_p$ | $t_r$ |
|---|---|---|---|---|---|---|---|---|
| IAE | 0.2721 | 3.9607 | 4.5915 | 3.8726 | 0.9996 | 0.5712 | 1.6 | 0.33 |
| ISE | 0.0101 | 28.9897 | 32.6211 | 22.3986 | 0.3330 | 1.6265 | 4.5 | 0.02 |
| ITAE | 0.0904 | 20.2502 | 10.5056 | 2.2081 | 1.0002 | 1.4431 | 2.2 | 0.59 |
| ITSE | 0.0346 | 2.9201 | 3.1423 | 1.4296 | 1.0991 | 0.4489 | 2.2 | 1.96 |
| ISTES | 2.9197 | 1.3843 | 1.1917 | 0.8663 | 0.9930 | 0.9486 | 1.0 | 4.37 |
| ISTSE | 0.0544 | 1.1383 | 1.7583 | 0.7735 | 1.0059 | 0.2773 | 1.1 | 2.64 |
| Sum of all PIs ($J_{PI\_all}$) | 0.4637 | 35.9099 | 35.8308 | 4.195 | 1.4572 | 1.2517 | 1.8 | 0.1 |

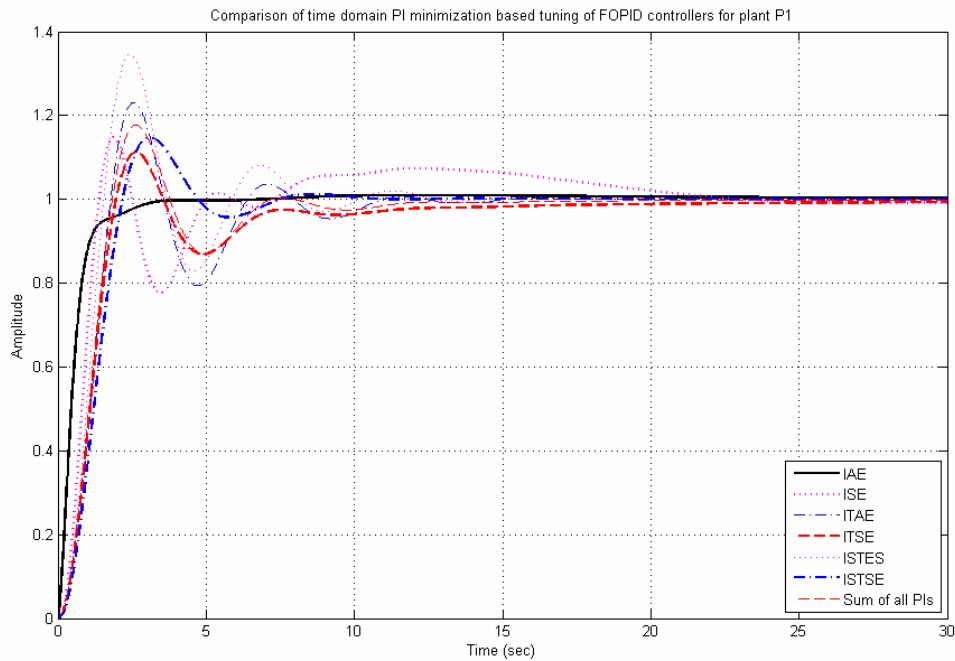

Fig. 3. Optimal performance index based tuning of FOPID controllers for plant $P_1$

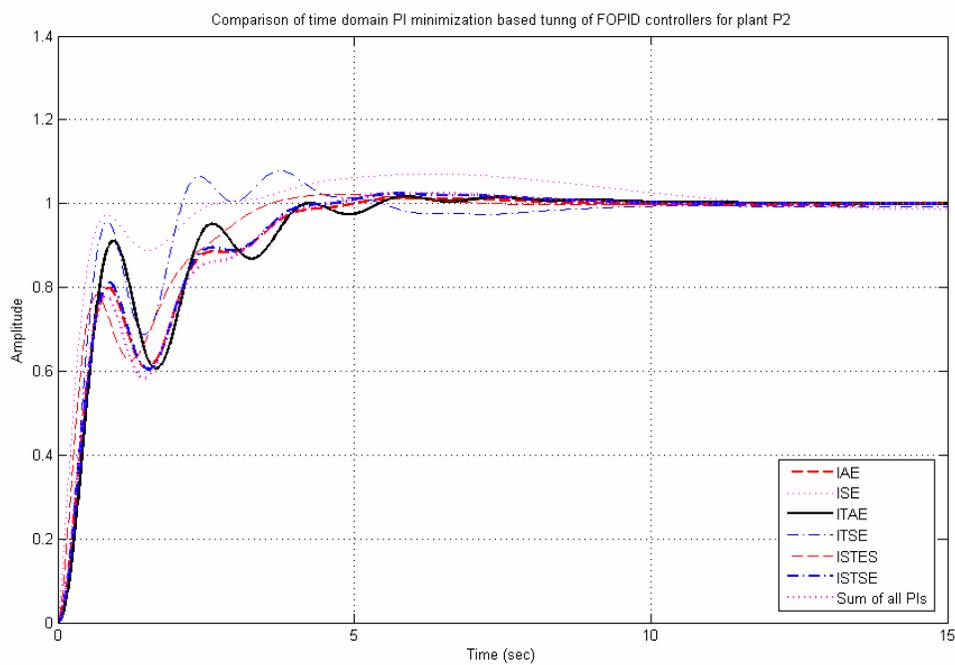

Fig. 4. Optimal performance index based tuning of FOPID controllers for plant $P_2$

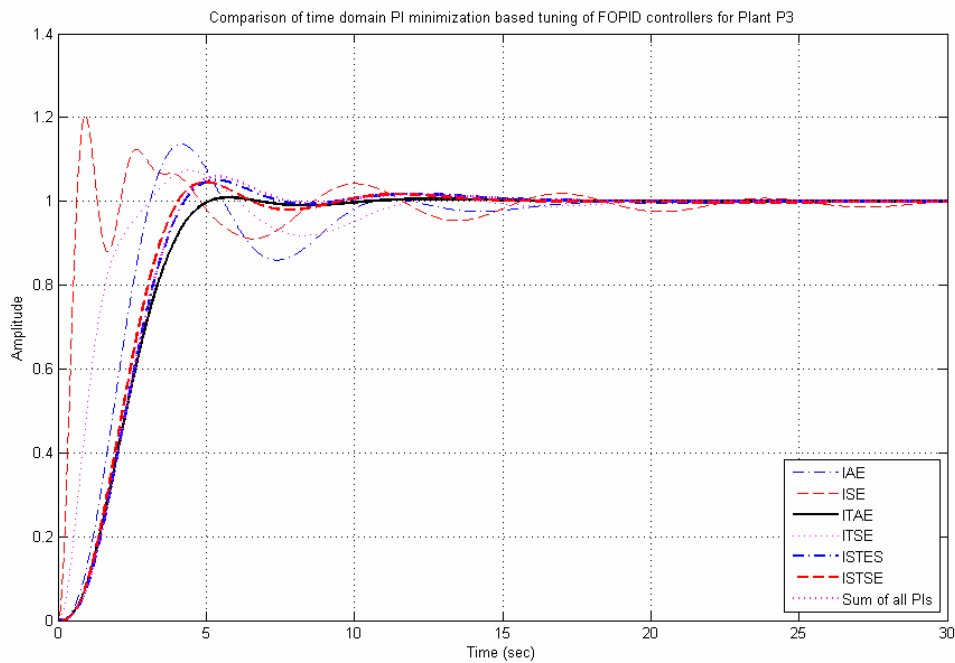

Fig. 5. Optimal performance index based tuning of FOPID controllers for plant $P_3$

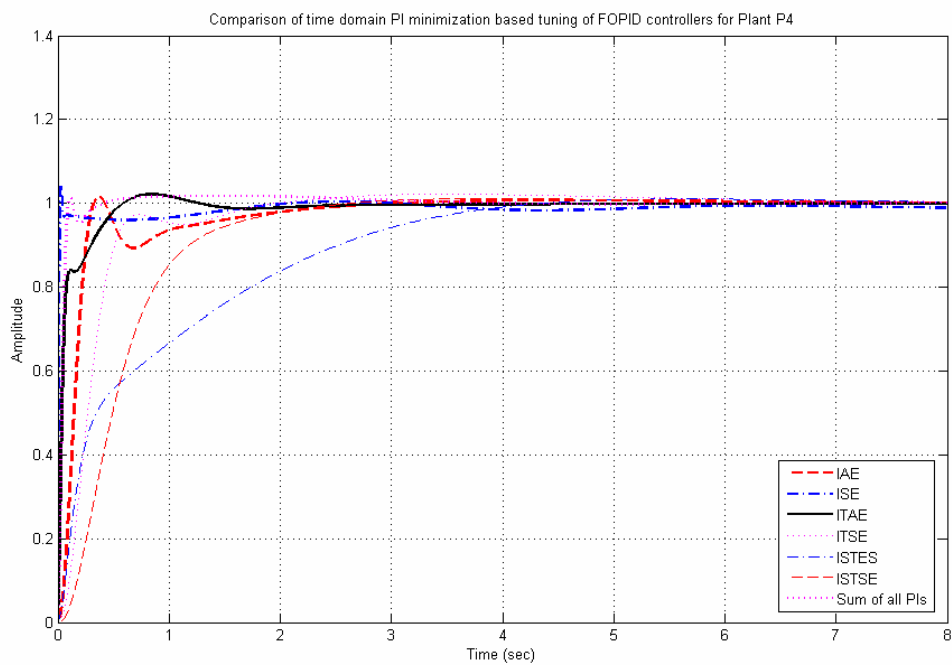

Fig. 6. Optimal performance index based tuning of FOPID controllers for plant $P_4$.

From Fig. 3-6, it can be concluded that in general, among all the integral performance indices, the ITAE criteria for FOPID controller tuning is capable of providing closed loop response with low overshoot and fast response, similar to that reported by Zhuang &

Atherton [55] for integer order PID controllers. Though for plant $P_1$ (Fig. 3), IAE has been found to be the best performance index over the others. So, for practical application on different higher processes, a comparison of different performance indices and a systematic engineering decision on the $\%M_p$ and $t_r$ are needed, as presented in this section. Infact, optimal tuning parameters with the most suitable performance index for a specific process may not produce optimal performance for other processess and hence the choice of performance index greatly depends on the process model itself for FOPID tuning and should not be chosen a priori.

## 5. Performance comparison of frequency and time domain design approaches:
### 5.1. Comparative results of parametric robustness (iso-damping property):

In section 3.3, the iso-damping nature of frequency domain design of FOPID controllers have been shown which uses a flat-phase criterion around $\omega_{gc}$ for controller tuning. On the other hand, the optimal time domain tuning presented in section 4.2 can not force the phase curve of the open loop system (comprising of the FOPID and the process plant) to be flat around $\omega_{gc}$. Hence robustness (in terms of the same $\%M_p$) can not be guaranteed for same amount of increase in loop gain, as reported in section 3.3. This fact is evident from the increase in overshoot with variation in system gain (Fig. 7) for time domain optimal tuning of FOPID controllers.

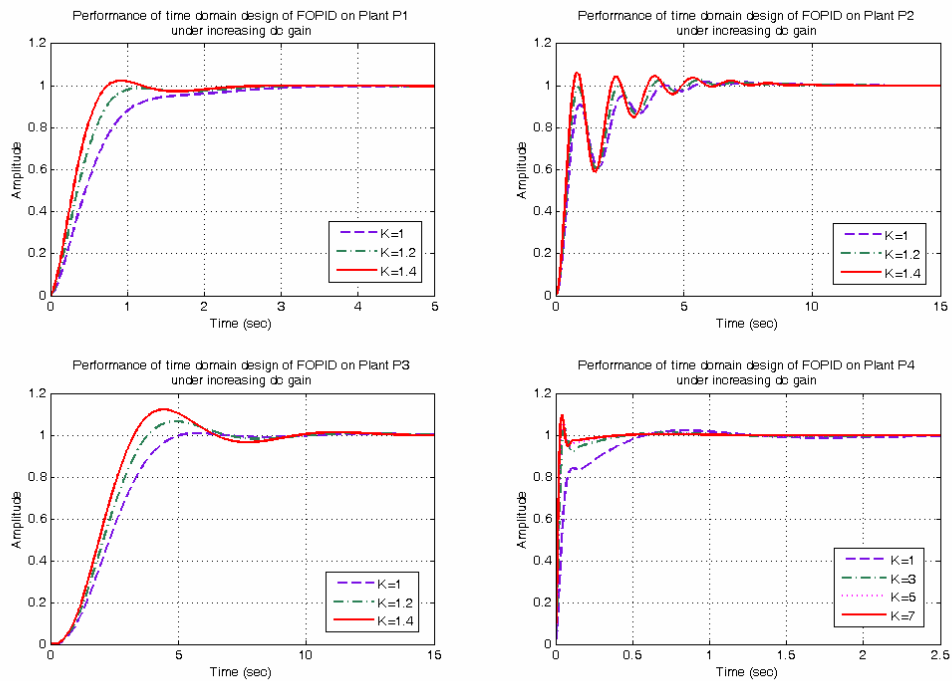

Fig. 7. Loss of iso-damping in time-domain performance index based optimal tuning.

The frequency domain design method, presented in section 3.3 uses an inherent robustness criterion while finding the controller parameters. This allows considerable variation in system gain to have a faster time response while keeping the overshoot constant (Fig. 1). In Fig. 7, controller parameters corresponding to the best response with minimum overshoot and fastest response in Fig. 3-6 are considered as the nominal cases (i.e. $K = 1$) for comparison.

### *5.2. Comparison of control signal and load disturbance rejection capability:*

It is well known that, the sensitivity function indicates the ability of the system to suppress load disturbances and achieve good set-point tracking. Whereas, the complementary sensitivity function indicates the robustness against measurement noise and other unmodelled system dynamics [19], [22]. To obtain a satisfactory time response under these disturbed conditions, the sensitivity function should have small values at lower frequencies and complementary sensitivity function should have small values at higher frequencies [26], [30]. Here, the magnitudes of sensitivity ($\left|S(s)\right|$) and complementary sensitivity ($\left|T(s)\right|$) are shown with the higher order process (12)-(15) and for both type of design of FOPID controllers (i.e. frequency and time domain) and compared in Fig. 8.

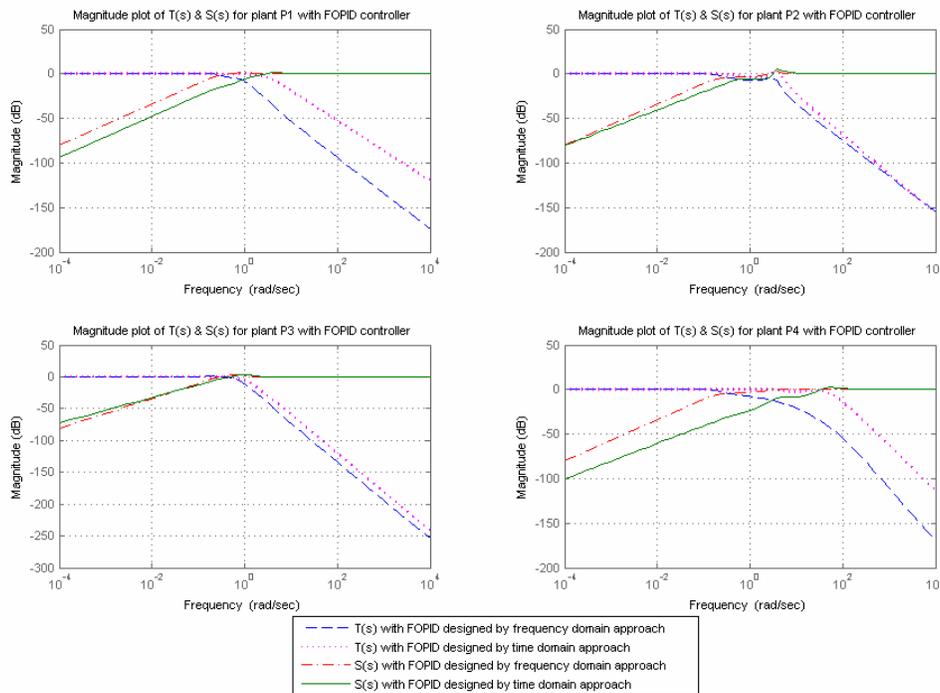

Fig. 8. Comparison of $\left|S(s)\right|$ and $\left|T(s)\right|$ with frequency and time domain design.

From Fig. 8, it is clear that the time domain design methodology of FOPID controllers has lower magnitude of sensitivity function at low frequencies. Hence it has better capability to suppress load disturbances. But the frequency domain design of FOPID is able to attenuate high frequency noises much better than that with the time domain design since the complementary sensitivity magnitudes are lower with it at higher frequencies. The best controllers obtained from time domain and frequency domain techniques are now tested with load disturbance (Fig. 9) which can also be predicted from the magnitude of sensitivity values itself in Fig 8. The corresponding control signals are also compared in Fig. 10. Clearly, lower value of control signal helps to reduce the size of the actuator and hence the cost involved and also the chance of actuator saturation and integral wind-up [19].

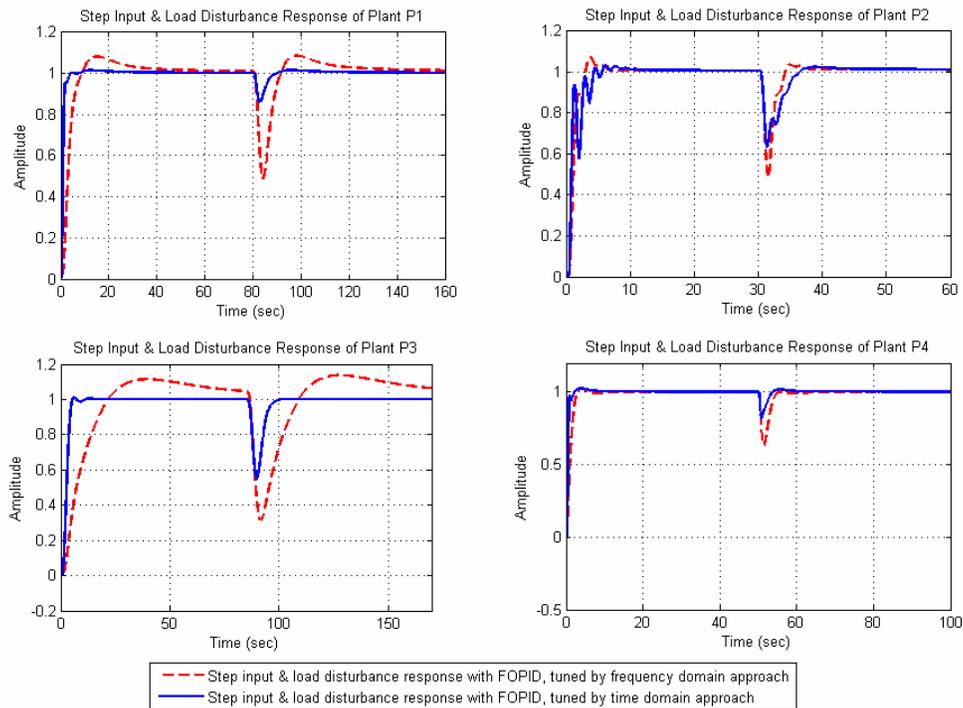

Fig. 9. Comparison of responses due to unit step change in set-point and load disturbance.

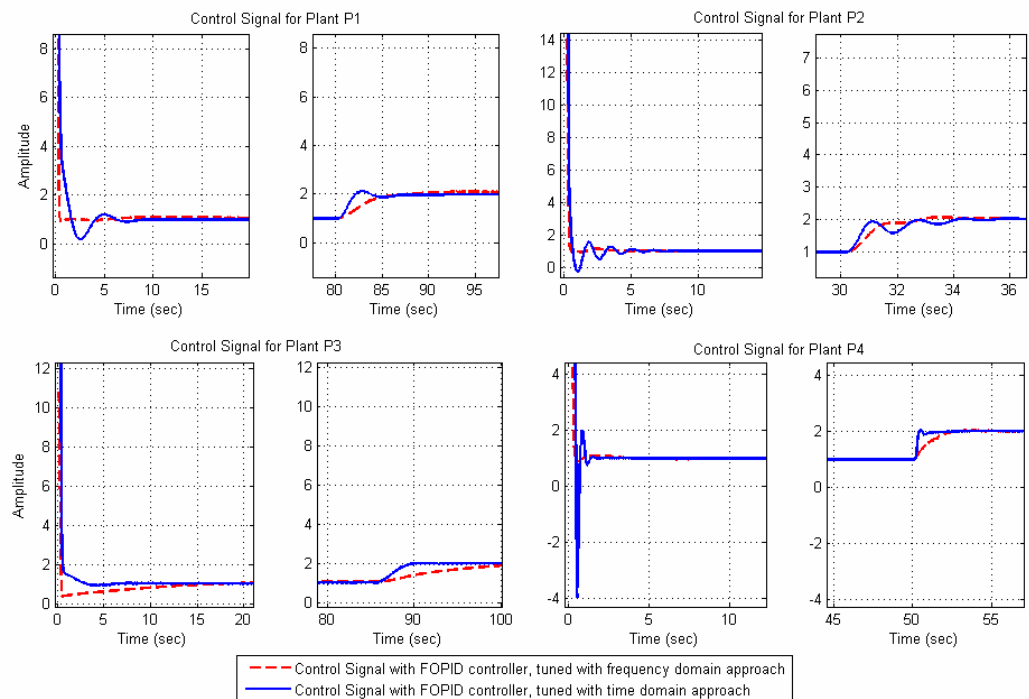

Fig. 10. Comparison of the control signals for the frequency and time domain.

Clearly in frequency domain design of FOPID controller, the load disturbance response is slightly poor in comparison with that with the time domain design (Fig. 9) but significant reduction in controller output signal is evident from Fig. 10. Thus it is evident that a single tuning technique cannot fulfill all of the contradictory controller design objectives simultaneously. Hence, selection of the tuning strategy can be done, depending on the nature of application in process industries and the priorities of design specifications to be met by the FOPID controller.

### 5.3. Summary of the results and few discussions

In the previous subsections, a comparative study on the frequency and time domain design of FOPID controllers are presented. It is shown that the frequency domain design methodology is capable of providing high robustness against loop gain variation but it can not be applied to a higher order process model directly. Hence it needs a reduced order modelling in some standard structures, among which NIOPTD-II has been found to be the most accurate one due to its superb flexibility to lower the modeling error (Table 1). Whereas, the time domain performance index minimization method does not require any additional model reduction technique and hence involves lesser computational load but with this methodology robustness can not be guaranteed. With parametric variation of the test plants, the performance of the controlled system deteriorates severely for time domain tuning of FOPID controllers. So, for offline tuning of FO-controllers a frequency domain method is always preferred where increased computational cost due to an extra model reduction technique involved, is not of a major concern. But for online controller tuning, having the process model well known from the governing system physical laws or system identification techniques, a time domain method can be easily applied since tuning of the controllers can be done much faster. Also, the time domain technique is capable of suppressing load disturbances much efficiently but on the other hand suffers from high frequency measurement noises and also higher chance of actuator saturation due to large control signal. So, for the time domain tuning technique, the implementation cost will be increased due to additional requirement of filters to attenuate high frequency measurement noises and also due to the large size of the actuator.

Also, in recent process control applications, many stochastic optimization algorithms is becoming popular for tuning of FOPID controllers namely PSO [33], [34], [36], [39], DE [32], GA [35], SOMA [31], IEMGA [40]. These intelligent optimization algorithms have been proved to give better performance over the deterministic optimization algorithms as these are able to take care of the trapping of the search at local minimas. But these stochastic algorithms take much computation time and also due to their randomness, satisfactory performance can not be guaranted without running the algorithms for a large number of times. Whereas a simple deterministic approach of optimization called Nelder-Mead Simplex algorithm [49], [50] with perturbed initial guesses (for time domain optimal tuning) or simultaneous nonlinear equation solving with Powell's Trust-Region-Dogleg algorithm (for robust frequency domain tuning) is capable of producing fairly accuarate model reduction and satisfactory controller design, with considerably faster and guaranted convergence with the proposed restrictions. Thus, the methodology presented in this paper is especially suitable for online parameter reduction of higher order processes and adaptive tuning of FOPID controllers, where stability guaranteed convergence and lesser computational load is of major concern.

Also, it is well known that FO elements are infinite dimensional linear filters [24] and hence creates a big problem in hardware realization. Practical implementation of FOPID controllers can be done by fractance and analog electronic circuit realization [24], [56]-[59], FPGA based digital realization [60] or electrochemical realization by lossy capacitors [24], [61], [62].

## 6. Conclusion

Comparative performance study of two design methodologies of FOPID controllers is done in this paper. The frequency domain approach is shown to give better performance in terms of robustness (iso-damping), better capability of high frequency noise rejection, lower value of control signal and hence reduced size of the actuator. The time domain optimal tuning methodology is faster but has lesser robustness but it has a nice ability to suppress load disturbances. On the other hand, it can't filter high frequency noises as efficiently as with the frequency domain tuning and also the control signal becomes very large which may saturate the actuator causing integral wind-up. From the above discussion, it can be concluded that no tuning methodology for a FOPID controller is *unconditionally beneficial*. Rather all the philosophies of controller tuning, discussed in this paper possess some strength and also some weakness and needs an engineering decision, depending on the nature of application in process controls.

The contributions of this paper can be summarized as:

- Proposal for two new templates for model parameter reduction namely NIOPTD-I and NIOPTD-II to obtain low modeling error than their conventional integer order counterparts i.e. FOPTD and SOPTD.
- Enhancement of robustness of a FOPID controller for frequency domain tuning technique with highly accurate (*flexible order*) reduced parameter templates.
- The available frequency domain robust FOPID design methodology is modified from a constrained optimization problem to a simultaneous nonlinear equation solving problem which takes lesser computational load and lesser complexity.
- Time domain tuning of FOPID controllers is enhanced with various integral performance indices while also choosing the most suitable performance criteria for optimization. Stability preserved tuning is also guaranteed with additional constraints imposed in the optimization process.
- A brief comparison of control performance (e.g. measurement noise filtration, small control signal and actuator size, load disturbance rejection) for time domain and frequency domain tuning technique are presented in this paper, along with few recommendations regarding their practical applicability like online and offline tuning.

Future scope of work can be directed towards fractional order modelling of open loop unstable plants; plants with fractional differ-integrators with several minimum or non-minimum-phase zeros and desgning suitable frational order controllers for such processes.

## ACKNOWLEDGEMENT

This work has been supported by the Department of Science & Technology (DST), Govt. of India under the PURSE programme.

**Reference:**

[1] Alf J. Isaksson and Stefan F. Graebe, "Analytical PID parameter expressions for higher order systems", *Automatica*, Volume 35, Issue 6, pp. 1121-1130, June 1999.

[2] G.M. Malwatkar, S.H. Sonawane and L.M. Waghmare, "Tuning PID controllers for higher-order oscillatory systems with improved performance", *ISA Transactions*, Volume 48, Issue 3, pp. 347-353, July 2009.

[3] Dale E. Seborg, Thomas F. Edgar and Duncan A. Mellichamp, "Process dynamics and control", John Wiley & Sons, New Delhi, 2009.

[4] Suman Saha, Saptarshi Das, Ratna Ghosh, Bhaswati Goswami, R. Balasubramanian, A.K. Chandra, Shantanu Das and Amitava Gupta, "Design of a fractional order phase shaper for iso-damped control of a PHWR under step-back condition", *IEEE Transactions on Nuclear Science*, Volume 57, Issue 3, pp. 1602-1612, June 2010.

[5] Karl J. Astrom and Tore Hagglund, "PID controller: Theory, design and tuning", Instrument Society of America, 1995.

[6] Aidan O' Dwyer, "Handbook of PI and PID controller tuning rules", Imperial College Press, London, U.K., 2006.

[7] Igor Podlubny, "Fractional-order systems and $PI^{\lambda}D^{\mu}$ controllers", *IEEE Transactions on Automatic Control*, Volume 44, Issue 1, pp. 208-214, January 1999.

[8] Ivo Petras, Lubomir Dorcak and Imrich Kostial, "Control quality enhancement by fractional order controllers", *Acta Montanistica Slovaca*, Volume 3, Number 2, pp. 143-148, 1998.

[9] Khalfa Bettou and Abdelfatah Charef, "Control quality enhancement using fractional $PI^{\lambda}D^{\mu}$ controller", *International Journal of Systems Science*, Volume 40, Issue 8, pp. 875-888, August 2009.

[10] Dingyu Xue and YangQuan Chen, "Sub-optimum $H_2$ rational approximations to fractional order linear systems", *in ASME 2005 International Design Engineering Technical Conferences and Computers and Information in Engineering Conference, IDETC/CIE 2005*, pp. 1527-1536, 24-28 September, 2005, Long Beach, California, USA.

[11] Dingyu Xue and YangQuan Chen, "A comparative introduction of four fractional order controllers", *in Proceedings of the 4th World Congress on Intelligent Control and Automation*, Volume 4, pp. 3228-3235, 10-14 June 2002, Shanghai, China.

[12] Alain Oustaloup and Pierre Melchior, "The great principles of the CRONE control", *in International Conference on Systems, Man and Cybernetics, 1993, 'Systems Engineering in the Service of Humans'*, Volume 2, pp. 118-129, 17-20 October 1993, Le Touquet.

[13] A. Oustaloup and M. Bansard, "First generation CRONE control", *in International Conference on Systems, Man and Cybernetics, 1993, 'Systems Engineering in the Service of Humans'*, Volume 2, pp. 130-135, 17-20 October 1993, Le Touquet.

[14] Alain Oustaloup, Benoit Mathieu and Patrick Lanusse, "Second generation CRONE control", *in International Conference on Systems, Man and Cybernetics, 1993, 'Systems Engineering in the Service of Humans'*, Volume 2, pp. 136-142, 17-20 October 1993, Le Touquet.

[15] Patrick Lanusse, Alain Oustaloup and Benoit Mathieu, "Third generation CRONE control", *in International Conference on Systems, Man and Cybernetics, 1993, 'Systems Engineering in the Service of Humans'*, Volume 2, pp. 149-155, 17-20 October 1993, Le Touquet.

[16] Concepcion A. Monje, Antonio J. Caderon, Blas M. Vinagre and Vicente Feliu, "The fractional order lead compensator", *in Second IEEE International Conference on Computational Cybernetics, ICCC 2004*, pp. 347-352, Vienna.

[17] YangQuan Chen, Kevin L. Moore, Blas M. Vinagre and Igor Podlubny, "Robust PID controller autotuning with a phase shaper", *in Proceedings of the first IFAC Symposium on Fractional Differentiation and its Application (FDA04)*, Bordeaux, France, 2004.

[18] Suman Saha, Saptarshi Das, Ratna Ghosh, Bhaswati Goswami, Amitava Gupta, R. Balasubramanian, A.K. Chandra and Shantanu Das, "Fractional order phase shaper design with Routh's criterion for iso-damped control system", *in 2009 Annual IEEE India Conference (INDICON)*, pp. 1-4, 18-20 December 2009, Gujarat.

[19] Suman Saha, Saptarshi Das, Ratna Ghosh, Bhaswati Goswami, R. Balasubramanian, A.K. Chandra, Shantanu Das and Amitava Gupta, "Fractional order phase shaper design with Bode's integral for iso-damped control system", *ISA Transactions*, Volume 49, Issue 2, pp. 196-206, April 2010.

[20] Duarte Valerio and Jose Sa da Costa, "Tuning of fractional PID controllers with Ziegler-Nichols-type rules", *Signal Processing*, Volume 86, Issue 10, pp. 2771-2784, October 2006.

[21] K. J. Astrom and T. Hagglund, "Revisiting the Ziegler-Nichols step response method for PID control", *Journal of Process Control*, Volume 14, Issue 6, pp. 635-650, September 2004.

[22] YangQuan Chen, Tripti Bhaskaran, Dingyu Xue, "Practical tuning rule development for fractional order proportional and integral controllers", *ASME Journal of Computational and Nonlinear Dynamics*, Volume 3, Issue 2, 021403, April 2008.

[23] Weng Khuen Ho, Chang Chieh Hang and Lisheng S. Cao, "Tuning of PID controllers based on gain and phase margin specifications", *Automatica*, Volume 31, Issue 3, pp. 497-502, March 1995.

[24] Shantanu Das, "Functional fractional calculus for system identification and controls", Springer, Berlin, 2008.

[25] YangQuan Chen, Kevin L. Moore, "Relay feedback tuning of robust PID controllers with iso-damping property", *IEEE Transactions on Systems, Man and Cybernetics-Part B: Cybernetics*, Volume 35, Issue 1, pp. 23-31, February 2005.

[26] Concepcion A. Monje, Antonio J. Calderon, Blas M. Vingre, YangQuan Chen and Vicente Feliu, "On fractional $PI^{\lambda}$ controllers: some tuning rules for robustness to plant uncertainties", *Nonlinear Dynamics*, Volume 38, Numbers 1-2, pp. 369-381, December 2004.

[27] HongSheng Li, Ying Luo and YangQuan Chen, "A fractional order proportional and derivative (FOPD) motion controller: tuning rule and experiments", *IEEE Transactions on Control System Technology*, Volume 18, Issue 2, pp. 516-520, March 2010.

[28] Ying Luo and YangQuan Chen, "Fractional order [proportional derivative] controller for a class of fractional order systems", *Automatica*, Volume 45, Issue 10, pp. 2446-2450, October 2009.

[29] Ying Luo, Yang Quan Chen, Chun Yang Wang and You Guo Pi, "Tuning fractional order proportional integral controllers for fractional order systems", *Journal of Process Control*, Volume 20, Issue 7, pp. 823-831, August 2010.

[30] Concepcion A. Monje, Blas M. Vinagre, Vicente Feliu and YangQuan Chen, "Tuning and auto-tuning of fractional order controllers for industry applications", *Control Engineering Practice*, Volume 16, Issue 7, pp. 798-812, July 2008.

[31] Lubomir Dorcak, Jan Terpak, Marcela Papajova, Frantiska Dorcakova and Ladislav Pivka, "Design of the fractional-order $PI^\lambda D^\mu$ controller based on the optimization with self-organizing migrating algorithm", *Acta Montanistica Slovaca*, Volume 12, Number 4, pp. 285-293, 2007.

[32] Arijit Biswas, Swagatam Das, Ajith Abraham and Sambarta Dasgupta, "Design of fractional-order $PI^\lambda D^\mu$ controllers with an improved differential evolution", *Engineering Applications of Artificial Intelligence*, Volume 22, Issue 2, pp. 343-350, March 2009.

[33] Deepyaman Maiti, Mithun Chakraborty, Ayan Acharya and Amit Konar, "Design of a fractional-order self-tuning regulator using optimization algorithms", *in Proceedings of 11th International Conference on Computer and Information Technology, ICCIT 2008*, pp. 470-475, 24-27 December 2008, Khulna, Bangladesh.

[34] Jun-Yi Cao and Bing-Gang Cao, "Design of fractional order controllers based on particle swarm optimization", *International Journal of Control, Automation and Systems*, Volume 4, Issue 6, pp. 775-781, December 2006.

[35] Jun-Yi Cao and Bing-Gang Cao, "Optimization of fractional order PID controllers based on genetic algorithm", *in Proceedings of the International Conference on Machine Learning and Cybernetics, ICMLC 2005*, pp. 5686-5689, Guangzhou, 18-21 August 2005.

[36] Deepyaman Maiti, Ayan Acharya, Amit Konar and Ramadoss Janarthanan, "Tuning PID and $PI^\lambda D^\mu$ controllers using the integral time absolute error criterion", *in 4th International Conference on Information and Automation for sustainability, ICIAFS 2008*, pp. 457-462, 12-14 December 2008, Colombo.

[37] Muwaffaq Irsheid Alomoush, "Load frequency control and automatic generation control using fractional-order controllers", *Electrical Engineering*, Volume 91, Number 7, pp. 357-368, March 2010.

[38] Mohammad Saleh Tavazoei, "Notes on integral performance indices in fractional-order control systems", *Journal of Process Control*, Volume 20, Issue 3, pp. 285-291, March 2010.

[39] Majid Zamani, Masoud Karimi-Ghartemani, Nasser Sadati and Mostafa Parniani, "Design of a fractional order PID controller for an AVR using particle swarm optimization", *Control Engineering Practice*, Volume 17, Issue 12, pp. 1380-1387, December 2009.

[40] Ching-Hung Lee and Fu-kai Chang, "Fractional-order PID controller optimization via improved electromagnetism-like algorithm", *Expert Systems with Applications*, Volume 37, Issue 12, pp. 8871-8878, December 2010.

[41] Mohamed Karim Bouafoura and Naceur Benhadj Braiek, "$PI^\lambda D^\mu$ controller design for integer and fractional plants using piecewise orthogonal functions", *Communications in Nonlinear Science and Numerical Simulation*, Volume 15, Issue 5, pp. 1267-1278, May 2010.

[42] Fernando J. Castillo, V. Feliu, R. Rivas and L. Sanchez, "Design of a class of fractional controllers from frequency specifications with guaranteed time domain behavior", *Computers and Mathematics with Applications*, Volume 59, Issue 5, pp. 1656-1666, March 2010.

[43] Varsha Bhambhani, Yiding Han, Shayok Mukhopadhyay, Ying Luo and YangQuan Chen, "Hardware-in-the-loop experimental study on a fractional order networked control system testbed", *Communications in Nonlinear Science and Numerical Simulation*, Volume 15, Issue 9, pp. 2486-2496, September 2010.

[44] Riccardo Caponetto, Giovanni Dongola, Luigi Fortuna and Antonio Gallo, "New results on the synthesis of FO-PID controllers", *Communications in Nonlinear Science and Numerical Simulation*, Volume 15, Issue 4, pp. 997-1007, April 2010.

[45] Serdar Ethem Hamamci and Muhammet Koksal, "Calculation of all stabilizing fractional-order PD controllers for integrating time delay systems", *Computers and Mathematics with Applications*, Volume 59, Issue 5, pp. 1621-1629, March 2010.

[46] Serdar Ethem Hamamci, "An algorithm for stabilization of fractional-order time delay systems using fractional-order PID controllers", *IEEE Transactions on Automatic Control*, Volume 52, Issue 10, pp. 1964-1969, October 2007.

[47] F. Merrikh-Bayat and M. Karimi-Ghartemani, "Method for designing $PI^{\lambda}D^{\mu}$ stabilisers for minimum-phase fractional-order systems", *IET Control Theory & Applications*, Volume 4, Issue 1, pp. 61-70, January 2010.

[48] Fabrizio Padula and Antonio Visioli, "Tuning rules for optimal PID and fractional-order PID controllers", *Journal of Process Control*, (In Press).

[49] J. A. Nelder and R. Mead, "A simplex method for function minimization", *The Computer Journal*, Volume 7, Issue 4, pp. 308-313, 1965.

[50] "MATLAB optimization toolbox: user's guide", Mathworks, Inc., 2009.

[51] K. J. Astrom, H. Panagopoulos and T. Hagglund, "Design of PI controllers based on non-convex optimization", *Automatica*, Volume 34, Issue 5, pp. 585-601, May 1998.

[52] H. Panagopoulos, K. J. Astrom and T. Hagglund, "Design of PID controllers based on constrained optimisation", *IEE Proceedings-Control Theory and Applications*, Volume 149, Issue 1, pp. 32-40, January 2002.

[53] Jing-Chung Shen, "New tuning method for PID controller", *ISA Transactions*, Volume 41, Issue 4, pp. 473-484, October 2002.

[54] Haiyang Chao, Ying Luo, Long Di and YangQuan Chen, "Roll-channel fractional order controller design for a small fixed-wing unmanned aerial vehicle", *Control Engineering Practice*, Volume 18, Issue 7, pp. 761-772, July 2010.

[55] M. Zhuang and D.P. Atherton, "Automatic tuning of optimum PID controllers", *IEE Proceedings D: Control Theory and Applications*, Volume 140, Issue 3, pp. 216-224, 1993.

[56] Gary W. Bohannan, "Analog fractional order controller in temperature and motor control applications", *Journal of Vibration and Control*, Volume 14, Issue 9-10, pp. 1487-1498, 2008.

[57] I. Podlubny, I. Petras, B.M. Vinagre, P. O'Leary and L. Dorcak, "Analogue realizations of fractional-order controllers", *Nonlinear Dynamics*, Volume 29, Numbers 1-4, pp. 281-296, July 2002.

[58] A. Charef, "Analogue realization of fractional-order integrator, differentiator and fractional $PI^{\lambda}D^{\mu}$ controller", *IEE Proceedings: Control Theory and Applications*, Volume 153, Issue 6, pp. 714-720, November 2006.

[59] Guillermo E. Santamaria, Jose V. Valverde, Raquel Perez-Aloe and Blas M. Vinagre, "Microelectronic implementations of fractional-order integrodifferential

operators", *ASME Journal of Computational and Nonlinear Dynamics*, Volume 3, Issue 2, pp. 021301, April 2008.

[60] Riccardo Caponetto and Giovanni Dongola, "Field programmable analog array implementation of noninteger order $PI^\lambda D^\mu$ controller", *ASME Journal of Computational and Nonlinear Dynamics*, Volume 3, Issue 2, pp. 021302, April 2008.

[61] Karabi Biswas, Siddhartha Sen and Pranab Kumar Dutta, "Realization of a constant phase element and its performance study in a differentiator circuit", *IEEE Transactions on Cicuits and Systems II: Express Briefs*, Volume 53, Issue 9, pp. 802-806, September 2006.

[62] Rodolfo Martin, Jose J. Quintana, Alejandro Ramos and Ignacio de la Nuez, "Modeling of electrochemical double layer capacitors by means of fractional impedance", *ASME Journal of Computational and Nonlinear Dynamics*, Volume 3, Issue 2, pp. 021303, April 2008.